\documentclass{aa}
\usepackage{graphicx}

\newcommand\teff{$ {\rm T_{eff}}$}

\newcommand\logg{$\log {\rm g}$}

\newcommand{\Msolar}{\mbox{\,$\rm M_{\odot}$}}        

\newcommand{\cii}{C {\sc ii}}
\newcommand{\ciii}{C {\sc iii}}
\newcommand{\nii}{N {\sc ii}}
\newcommand{\niii}{N {\sc iii}}
\newcommand{\oi}{O {\sc i}}
\newcommand{\oii}{O {\sc ii}}
\newcommand{\nei}{Ne {\sc i}}
\newcommand{\neii}{Ne {\sc ii}}
\newcommand{\mgii}{Mg {\sc ii}}
\newcommand{\alii}{Al {\sc ii}}
\newcommand{\aliii}{Al {\sc iii}}
\newcommand{\silii}{Si {\sc ii}}
\newcommand{\siliii}{Si {\sc iii}}
\newcommand{\siliv}{Si {\sc iv}}
\newcommand{\pii}{P {\sc ii}}
\newcommand{\piii}{P {\sc iii}}
\newcommand{\sii}{S {\sc ii}}
\newcommand{\siii}{S {\sc iii}}
\newcommand{\arii}{Ar {\sc ii}}
\newcommand{\titii}{Ti {\sc ii}}
\newcommand{\feii}{Fe {\sc ii}}
\newcommand{\feiii}{Fe {\sc iii}}

\begin{document}
\title{Early type stars at high galactic latitudes\\
       II. Four evolved B-type stars of unusual chemical composition
\thanks{Based on observations 
    obtained 
    at the W.M. Keck Observatory, which is operated by the Californian 
    Association for Research in Astronomy for the California Institute of 
    Technology and the University of California
}
\thanks{Based on observations collected at the German-Spanish Astronomical 
Center (DSAZ), Calar Alto, operated by the Max-Planck-Institut f\"ur 
Astronomie Heidelberg jointly with the Spanish National Commission for 
Astronomy} 
\thanks{Based on observations collected at the European Southern
        Observatory (ESO proposal No. 65.H-0341(A))}
}

\author{M. Ramspeck
 \and U. Heber
 \and H. Edelmann}

\offprints{M. Ramspeck}

\institute{Dr.-Remeis-Sternwarte, Universit\"at Erlangen-N\"urnberg,
           Sternwartstr. 7,
           D-96049 Bamberg, Germany
           e-mail: ramspeck@sternwarte.uni-erlangen.de}

\date{Received 26 July 2001/Accepted 10 September 2001}

\abstract{
We present the result of differential spectral analyses of a further four  
apparently normal B-type stars. 
Abundance anomalies (e.g. He, C, N enrichment), slow rotation and/or high gravities
suggest that 
the programme stars are evolved low-mass B-type stars. In order to trace 
their evolutionary 
status several scenarios are discussed. Post-AGB evolution can be ruled out.
PG~0229$+$064 and PG~1400$+$389 could be horizontal branch (HB) stars, 
while HD~76431 and SB~939 have already evolved away from the 
extreme HB (EHB). 
The low helium 
abundance of HD~76431 is consistent with post-EHB evolution. 
The enrichment in helium, carbon and nitrogen can be explained either by 
deep mixing of nuclearly processed material to the surface or by diffusion
processes modified by magnetic fields and/or stellar winds. A kinematic 
study of their galactic orbits indicates that the stars belong to an old 
disk population. 
\keywords{galaxy: halo -- stars: early-type -- stars: abundances -- stars: kinematics 
 -- stars: evolution}}
\authorrunning{Ramspeck et al.}
\titlerunning{Early type stars at high galactic latitudes II}
\maketitle

\section{Introduction\label{intro}} 
The population of faint blue stars at high galactic latitudes is dominated 
by subluminous O- and B-type stars. However, several objects among the faint 
blue stars are apparently normal B-type stars because their spectra closely
resemble that of main sequence stars (see Tobin \cite{tobi87}, 
Keenan \cite{keen92} and Heber et al., \cite{hemo97} for reviews).

Tobin (\cite{tobi87}) discusses the problem that some highly evolved
stars spectroscopically mimic normal massive stars almost perfectly. The
most striking example is PG\,0832+676 which has been analysed several
times. Its abundance pattern is close to normal. Only recently, Hambly et
al. (\cite{hake96}) were able to firmly establish slight underabundances and a very
low projected rotation velocity. Combining both results they
concluded that PG\,0832+676 in fact is a highly evolved
star. In a recent paper, Hambly et al. (\cite{haro97}) extended their study 
to a dozen apparently normal B-type stars and demonstrated that they are also 
of low-mass. 

Abundance analyses as well as determinations of rotational velocities
are thus essential to distinguish massive from low-mass stars.
A high rotational velocity generally excludes a late
evolutionary status of the star, as old, low-mass stars cannot rotate as fast
as massive stars.

During our ongoing investigation of faint blue stars we encountered several
apparently normal B-type stars (Moehler et al. 
\cite{mohe94}, Heber et al.~\cite{hemo95}, Schmidt et al.~\cite{scde96}). 
In a recent paper we described the 
spectral analysis of ten massive B-type stars at high galactic latitudes 
(Ramspeck et al. \cite{rahe01}, henceforth paper~I).

Here we present the analysis of new high-resolution spectra 
for four
additional stars which were classified as apparently normal from spectra
of lower spectral resolution.
We have obtained high resolution Echelle spectra for 
PG~0229$+$064, PG~1400$+$389, HD~76431 and SB~939 
using the HIRES spectrograph at the Keck~I telescope, the FOCES
spectrograph at the DSAZ 2.2m telescope and the CASPEC spectrograph at the ESO 
3.6m telescope.
The data set is supplemented by long slit spectra obtained at the DSAZ 3.5m
telescope and the ESO 1.5m telescope.
Details of the observations are given in Table~\ref{hlres} and the 
data reduction technique is outlined in paper~I. 

The atmospheric parameters, rotational 
velocities and metal abundances are determined in sections~\ref{param_section} 
and \ref{abu_section}. Then the evolutionary status 
(section~\ref{evol_section}) and 
the kinema\-tics of the stars (section~\ref{kine_section}) are discussed. 
The last 
section summarizes the conclusions.

\begin{table}
\caption[]{Observational details}
\label{hlres}
\begin{tabular}{|c|ccc|} \hline
Name & Obs. date & Tel. \&   &  $\Delta \lambda$ (nm)  \\ 
     &           &  Instrum. &    \\ \hline
     & \multicolumn{3}{c|}{Echelle spectra}  \\
 PG\,0229$+$064    & Jul 20, 1998 14:37 & K+H    & 360 -- 513  \\
                   & Sep 13, 1998 03:00 & CA+F   & 387 -- 683  \\
 PG\,1400$+$389    & Jul 23, 1996 20:00 & K+H    & 426 -- 670  \\
 HD\,76431         & Feb 01, 2000 02:00 & CA+F   & 389 -- 693  \\
 SB\,939           & Oct 1986           & E+C    & 406 -- 513  \\ 
\hline
                   & \multicolumn{3}{c|}{Long slit spectra}  \\
 PG\,0229$+$064 & Nov 21, 1988  04:39 & E+B\&C      &  403 -- 491\\
 PG\,1400$+$389 & Apr 14, 2001   --   & CA+T          & 410 -- 495\\
 HD\,76431      & Apr 13, 2001   --   & CA+T          & 410 -- 495\\
 \hline
\multicolumn{4}{l}{} \\
\multicolumn{4}{l}{K+H: KECK\,I + HIRES (spectral resolution: 0.09\AA)} \\
\multicolumn{4}{l}{CA+F: Calar Alto 2.2m + FOCES (sp. resolution: 0.15\AA)} \\
\multicolumn{4}{l}{E+C: ESO 3.6m + CASPEC (sp. resolution: 0.2\AA)} \\
\multicolumn{4}{l}{E+B\&C: ESO 1.5m + B\&C (sp. resolution: 2.5\AA)} \\
\multicolumn{4}{l}{CA+T: Calar Alto 3.5m + Twin (sp. resolution: 1.0\AA)} \\
\end{tabular}\\[2mm]
\end{table}

\section{Atmospheric Parameters and Rotational Velocities}\label{param_section}

Effective temperatures, surface gravities
and photospheric helium abundances are derived by matching the Balmer and 
helium line profiles to a grid of synthetic spectra calculated from LTE model 
atmospheres 
as described in detail in paper I.

HD~76431, however, turned out to be considerably hotter than the other 
programme stars and the assumption of LTE might be questionable.
Therefore a grid of partially 
line blanketed NLTE model atmospheres (Napiwotzki \cite{napi97}) was used. 
These models contain hydrogen and helium. The latest version of the NLTE 
code by Werner (\cite{wern86}) is used, which employs the Accelerated Lambda 
Iteration (ALI) technique (Werner \& Husfeld, \cite{wehu85}, see Werner \& 
Dreizler \cite{wedr99} for details).\\
\begin{figure}
\vspace{8.5cm}
\includegraphics{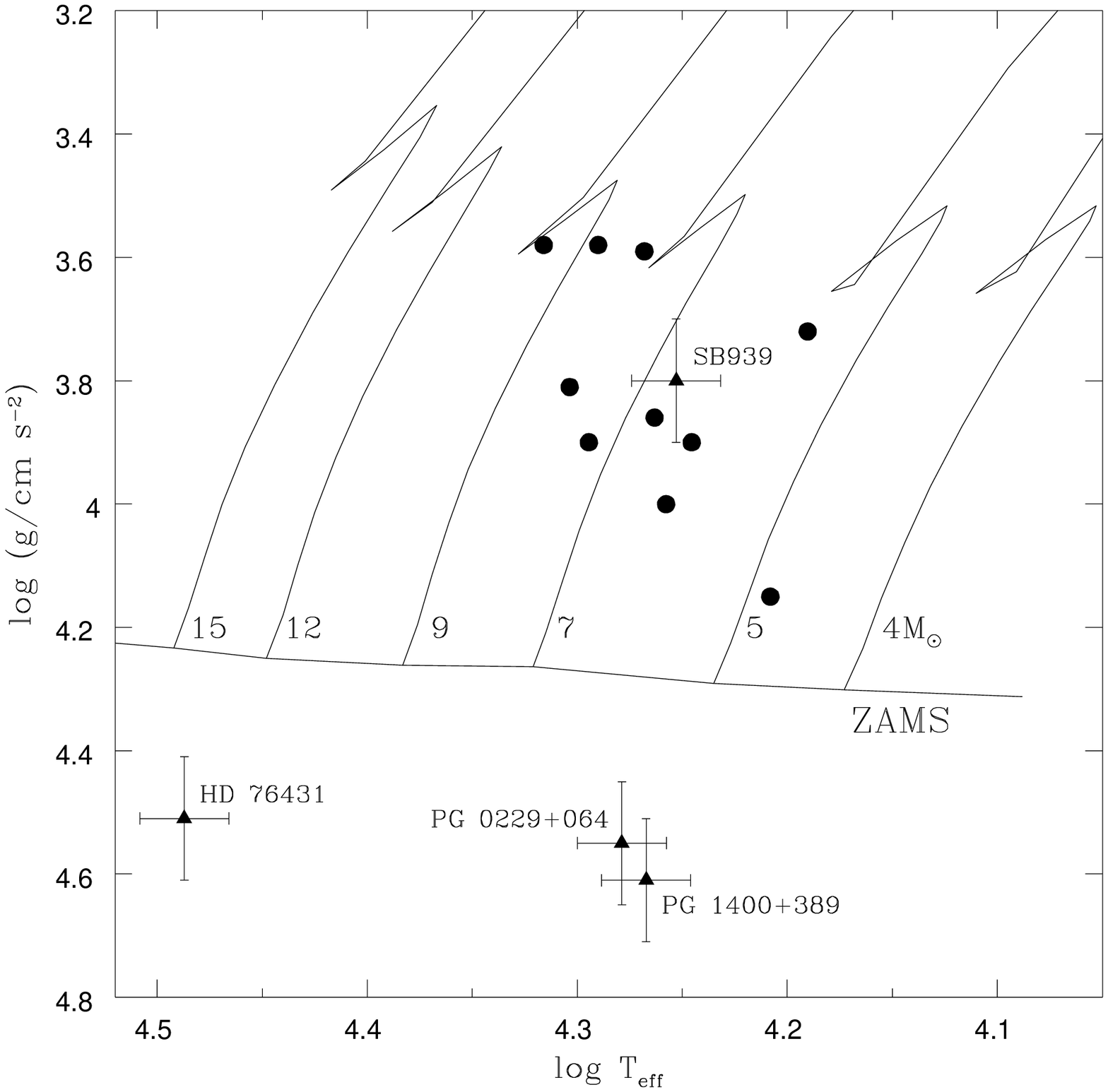}
 \caption[]{The positions of the programme stars (triangles with error bars) 
            and that of massive B-type stars
            from paper I (filled circles) in the (\teff, \logg) diagram are 
            compared 
            with evolutionary tracks calculated by Schaller et al.
            (1992) for main sequence stars.\\ The filled circles
            are objects with detectable rotation and the
            triangles for non-rotating objects (v\,sin\,i $<$5 km $\rm{s^{-1}}$).
\label{schaller}
}
\end{figure}
\begin{figure*}
\vspace{12.5cm}
\includegraphics{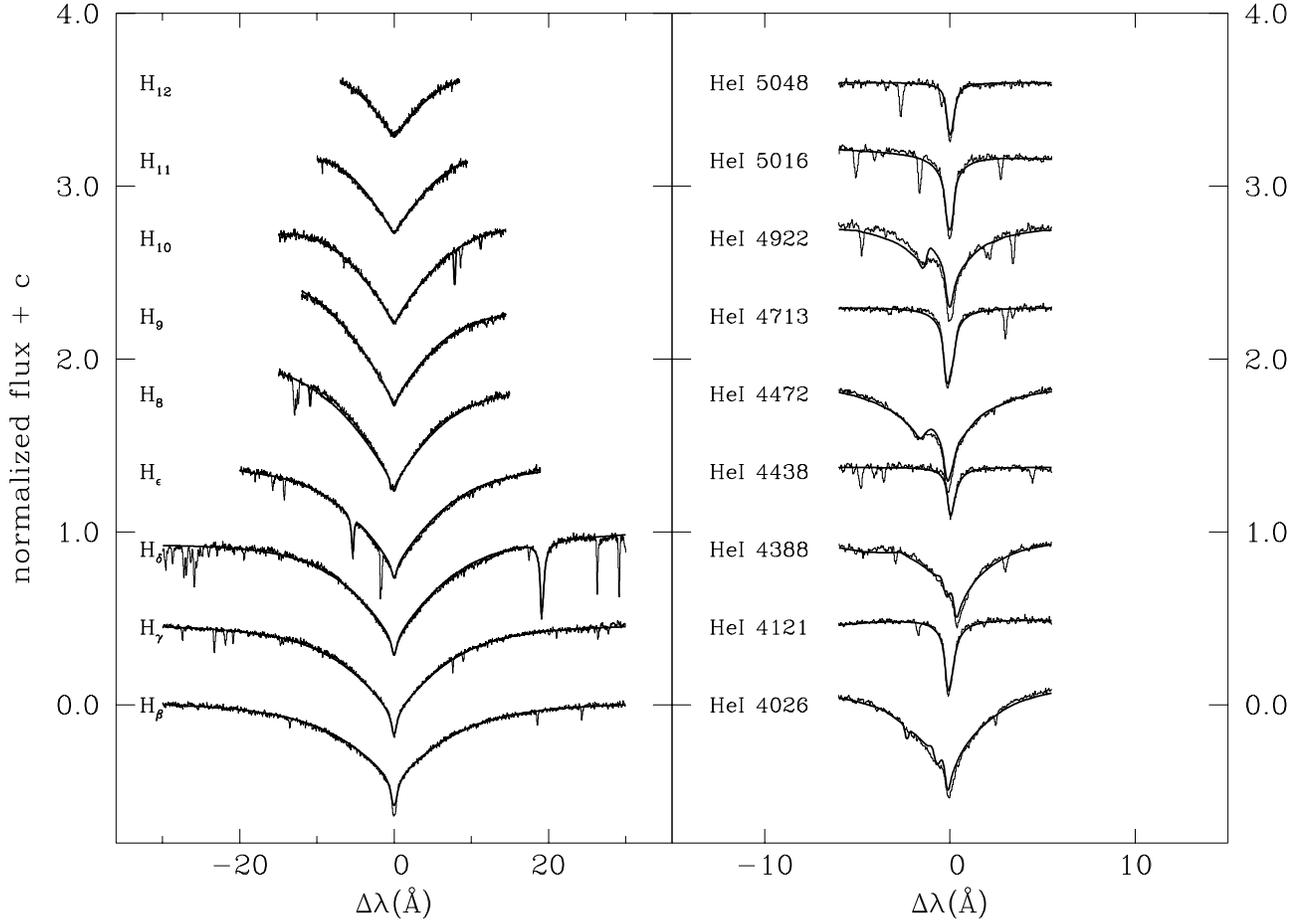}
\caption[]{Fit of the Balmer and helium lines by synthetic spectra 
for PG~0229$+$064.
 	\label{sp_fit}
}
\end{figure*}

\begin{table*}
\centering
\caption[]{Atmospheric parameters and rotational velocities for the programme
stars as derived from high and low resolution spectroscopic data and 
comparison of these data with the effective temperatures calculated from
Str\"omgren photometry. The upper limits of rotational velocities are determined from
the {\mgii} doublet at 4481\AA.}
\label{comparison}
\begin{tabular}{|c|cccc|ccc|cc|} \hline
      &   \multicolumn{4}{c|}{High Resolution} & \multicolumn{3}{c|}{Low 
Resolution}  & \multicolumn{2}{c|}{Photometry} \\
 Name & $\rm{T_{eff}}$  & $\rm{\log(\frac{g}{cm{  }s^{-2}})}$ & 
 $\rm{\log\frac{N(He)}{N(H)}}$ & $v \sin i$  &   
 $\rm{T_{eff}}$ & $\rm{\log(\frac{g}{cm{  }s^{-2}})}$ & 
 $\rm{\log\frac{N(He)}{N(H)}}$ &
 $\rm{T_{eff}}$ & E(b--y) \\ 
 & (K) &  &  &  (km $\rm{s^{-1}}$) & (K) & &  & (K) & \\ \hline
 PG\,0229$+$064 & 19\,000 & 4.55 & --0.80 & $<5$ & 19\,200 & 4.51 & --0.72 & 20\,200 (1) & 0.039 \\
               &         &      &       &       &         &      &       & 21\,000 (2)   & 0.036 \\
 PG\,1400$+$389 & 18\,200 & 4.51 & --0.60 & $<5$ & 18\,800 & 4.71 & --0.71 & 20\,500 (1) & 0.023 \\
                &         &      &        &      &         &      &        & 17\,100 (3) & 0.004 \\
 HD\,76431 & 31\,000 & 4.51 & --1.51 & $<5$ & 28\,500 & 4.31 & --1.68 &  30\,100     (4) & 0.0  \\
 SB\,939          & 17\,900 & 3.80 & --0.60 & $<5$ & -- & -- & -- & 18\,000 (5)          & 0.001 \\ \hline
\end{tabular} \label{fitres}\\[2mm]
\parbox[t]{150mm}{\it{Str{\o}mgren photometry from:}: 
(1) Wesemael et al. (\cite{wefo92}); (2) Moehler et al. (\cite{mori90});  
(3) Mooney et al. (\cite{moro00});
(4) Kilkenny (\cite{kil87}); (5) Graham J.A. et al. (\cite{grsl73})}
\end{table*}

\subsection{Effective temperatures and gravities}

The fit was executed for all high and low resolution spectra and the
results are listed in Table \ref{comparison}. 
An example fit is reproduced 
in Fig.~\ref{sp_fit} to illustrate the excellent quality of the matching of 
the observations by synthetic spectra.
Errors in effective temperatures were estimated
conservatively as 5\% and we adopted an error of $\pm$0.1 dex for the
gravities of all programme stars.

For all stars  Str\"omgren photometry 
is available, which allowed 
an independent determination of the effective temperature. We used 
the program of
Moon (\cite{moon85}) as modified by Napiwotzki et al. (\cite{nasc93}) to derive
the effective temperature and the reddening and 
compare the photometric temperatures to the spectroscopic ones in 
Table~\ref{comparison}.
There is a good agreement between results from low
and high resolution spectra and photometry, except for 
PG\,1400$+$389 for which the two available colour measurements 
result in rather different \teff.\\ 
The parameters used for further analyses were taken from the high resolution 
spectra, except for PG\,1400$+$389, because the spectral range covered is 
larger, allowing to use more Balmer and helium lines in the fit procedure.
However, in the case of PG~1400$+$389 the Echelle and the long slit spectrum 
covered only three Balmer lines. Therefore, we 
averaged  the parameter derived from high and low resolution spectra.

Our results are shown in a (\teff, \logg) diagram and compared to high 
latitude main sequence stars (from paper I) and to 
evolutionary tracks for main sequence stars 
(\cite{scsc92}, Fig. \ref{schaller}).
SB\,939 is located in the main sequence band. 
The other stars, however, lie below the
main sequence and therefore cannot be massive stars.
Further clues as to the nature of these stars come from the projected 
rotational velocities and the chemical abundance pattern.

\subsection{Helium abundances}

\begin{figure}
\vspace{12.5cm}
\includegraphics{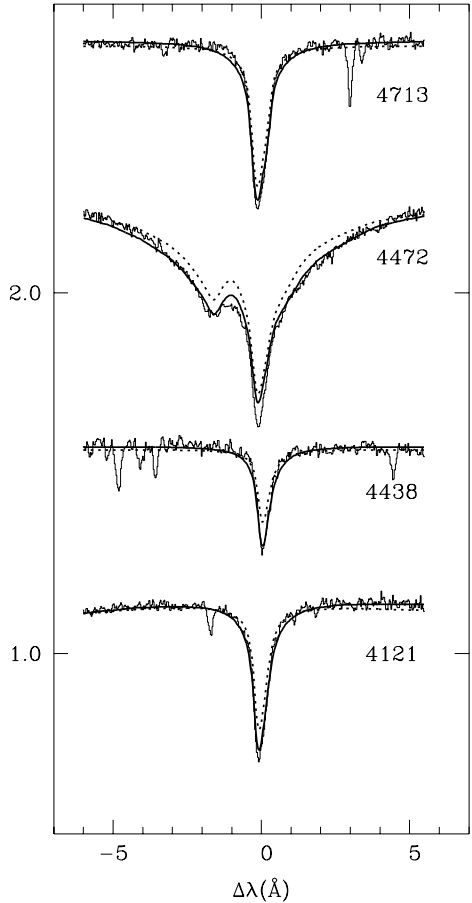}
\caption[]{Comparison of helium lines in the spectrum of PG~0229$+$064 to 
synthetic line profiles for helium abundance derived by the fit 
procedure and for normal helium abundance (dashed).
 	\label{helium}
}
\end{figure}

An outstanding result of the analysis is the non-solar helium abundance of 
all programme stars. 
While in HD~76431 helium is depleted by a factor of 3 with respect to 
the sun, helium is significantly enriched in the 
other stars.
Fig.~\ref{helium} illustrates the influence of the 
helium enrichment on the profiles of weak and strong He lines in the case 
of PG~0229$+$064, which is only mildly helium rich.

In previous spectral analyses of three of our programme stars
(Saffer et al. \cite{sake97})
based on low resolution spectra, the peculiar helium abundance could not be 
obtained
because the fit procedure was executed assuming a solar helium abundance.
This underestimate of the helium abundance also led to 
considerably higher effective temperatures and gravities for 
PG~0229+064 and PG~1400$+$389 than our analyses.

The non-solar helium abundances is another indicator that PG\,1400$+$389 
and PG\,0229$+$064 are 
unlikely to be normal massive B-type stars.\\
The high helium content of 
SB\,939 was already pointed out by Langhans  \& Heber (\cite{lahe86}) 
and is confirmed by our analysis. Langhans \& Heber (\cite{lahe86}) 
suggested that SB~939 might belong to the class of intermediate helium 
stars which are main sequence stars with unusually strong (and sometimes 
variable) helium lines (for a review see Hunger, \cite{hung86}). They 
populate a temperature range (22\,000\,K to 28\,000\,K) somewhat hotter
than SB~939 and are found close to the galactic plane
(Drilling \cite{dril86}). If confirmed,
SB~939 would be the first intermediate helium star at high galactic 
latitudes. But still we cannot exclude that SB~939 belongs to the class
of intermediate helium stars.

\subsection{Projected rotational velocities} 

All programme stars display very sharp absorption lines indicating that 
they are slow rotators unless they are seen pole-on. 
The components of the Mg~{\sc ii}  doublet (4481.13/4481.33\AA) are resolved
that can be used to estimate the projected rotational velocity. An upper 
limit of 5\,km\,s$^{-1}$ results for all stars.
These low values provide further evidence that the programme stars are 
unlikely to be main sequence stars.

\section{Metal Abundances}\label{abu_section}
The equivalent widths were measured employing the nonlinear least-squares
Gaussian fitting routines in MIDAS with central wavelength, central 
intensity and full width at half maximum as adjustable parameters.
For metal lines located in direct
neighbourhood of Balmer or helium lines an additional Lorentzian function is 
used to describe the line wings of the latter.

Metal lines of the species {\cii}, {\ciii}, {\nii}, {\niii}, {\oi}, {\oii},
{\nei}, {\neii}, {\mgii}, {\alii}, {\aliii}, {\silii}, {\siliii}, {\siliv},
{\pii}, {\piii}, {\sii}, {\siii}, {\arii}, {\titii}, {\feii} and {\feiii} 
could be 
identified. The atomic data for the analysis were taken from several tables: 
\begin{enumerate}
 \item CNO from Wiese et al. (\cite{wifu96})
 \item Fe from Kurucz (\cite{kuru92}) and Ekberg (\cite{ekbe93})
 \item Ne, Mg, Al, Si, S, P, Ar, Ti from Hirata et al. (\cite{hiho95})
\end{enumerate}

The LTE abundances were derived by using the classical curve--of--growth
method and the LINFOR program (version of Lemke, see paper I). 
In this case the model
atmospheres were generated for the appropriate values of effective
temperature, gravity and solar helium and metal abundance with the 
ATLAS9 program of Kurucz (\cite{kuru92}).\\ 

Then we calculated curves of growth for the observed metal lines, from which
abundances were derived. 
Blends from different ions were omitted from the analysis.
In the final step the abundances were determined from a detailed spectrum
synthesis (using LINFOR code described above) of all lines measured before.
The results of the LTE abundance analysis and the r.m.s. errors 
for PG\,0229$+$064, PG\,1400$+$389, SB\,939, 
and HD~76431 are shown in Table~\ref{LTE}. 
The number of lines used is given in brackets.
Beside the statistical r.m.s. errors 
the uncertainties in \teff, \logg{ }and microturbulent velocity (see below)
contribute to the error budget.

\begin{figure*}
\vspace{15.6cm}
\includegraphics{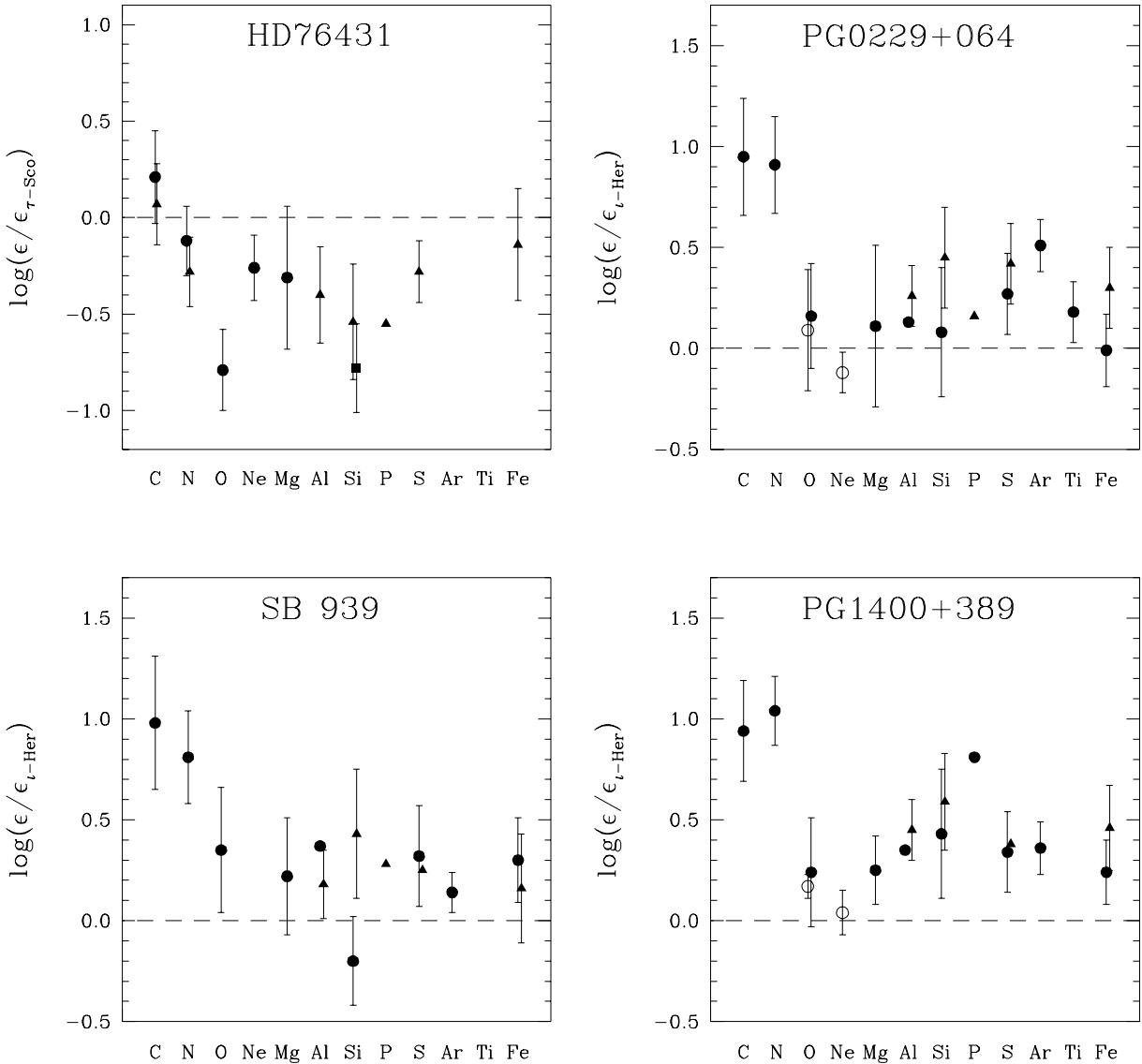}
\caption[]{LTE abundances (relative to $\tau$~Sco for HD~76431 and relative to $\iota$~Her
for the other three stars) and errors of the  programme stars. 
Abundances derived from neutral elements are shown as open circles, from
singly ionizied ones as filled circles, from doubly ionizied ones as filled 
triangles and from triply ionizied ones as filled squares. 
 \label{metal}
}
\end{figure*}

In order to minimize the systematic errors we have choosen the B-type star $\tau$~Sco
for HD~76431 and $\iota$~Her for the other three targets 
as comparison stars, since they have similar atmospheric parameters
to those of our programme stars. These stars have been ana\-ly\-sed by Hambly et al. (\cite{haro97}).
We redetermined the LTE abundances of $\tau$~Sco and $\iota$~Her using 
the same atomic data, model atmosphere and spectrum synthesis code as the
programme stars and took the equivalent widths of unblended lines 
measured by Hambly et al. (\cite{haro97}).\\  
Our results for $\iota$~Her agree to within 0.1~dex with those of Hambly et al. (\cite{haro97})
except for {\cii} (0.12~dex), {\siliii} (0.17~dex), {\siii} (0.21~dex) and {\feiii} (0.36~dex).
In particular our statistical error for {\feiii} is much lower than that of 
Hambly et al. (\cite{haro97}). \\
For $\tau$~Sco we also found good agreement with the results of Hambly et al. (\cite{haro97})
except for {\cii} (0.31~dex), {\ciii} (0.17~dex), {\niii} (0.13~dex), {\aliii} (0.18~dex),
{\siii} (0.17~dex) and {\feiii} (0.43~dex). 
The difference for $\iota$~Her and $\tau$~Sco between Hambly et al. (\cite{haro97})
and our work can be attributed to different oscillator strength used (esp. for {\feiii})
and our neglect of blended lines.
Results are given in Tables \ref{LTE} and \ref{HD76LTE} and systematic errors 
are adopted for our programme stars as well. 
These errors are incorporated in the error bars plotted in Fig. \ref{metal}.
The determination of elemental abundances is interlocked with the microturbulent
velocity $\xi$, that can be derived if a sufficient number of lines of one ion
can be measured over a wide range of line strengths. In our programme stars
{\nii} and {\oii} lines are most suitable for this purpose since 
many lines of these ions can be identified. Microturbulent velocities of 
$\xi$ = 5 km $\rm{s^{-1}}$ were found for PG~0229$+$064, 
PG~1400$+$389
and HD~76431, while a somewhat higher value of $\xi$~= 8~km $\rm{s^{-1}}$ was 
deduced for SB~939. 

In a differential analysis systematic errors should cancel to a large extent.
NLTE effects are small for all elements ($\le$0.1 dex, Kilian \cite{kili94}) 
except for {\nei}, which is overestimated by LTE 
calculations (Auer \& Mihalas, \cite{aumi73}) e.g. by 0.60~dex in the case 
of $\iota$~Her. Correcting our measurements for this NLTE effect the Ne 
abundances are quite close to that of normal B-type stars (Kilian \cite{kili94}).
Correcting for the NLTE effect on {\nei} 
(0.60~dex, see above) 
the neon abundance of the three
programme stars for which it has been measured is found 
to be consistent with Kilian's distribution of Ne abundances in normal B 
stars.

The abundances of programme stars with respect to 
$\tau$~Sco for HD~76431 and $\iota$~Her for the others
are plotted in Fig.~\ref{metal}.

It is well known that considerable variations of metal 
abundances from star to star occur among normal main sequence B-type stars (e.g.
Kilian, \cite{kili94}). This has to be taken into account when judging 
whether a measured abundance pattern is peculiar or not (see also paper I).
\begin{table*}
\centering
\caption[]{Mean absolute LTE abundances and r.m.s errors for programme stars. The numbers in brackets indicate
           the number of lines per ion. For $\iota$~Her the abundances are determined with the
	   equivalent widths from Hambly et al. (\cite{haro97}).
           Errors for the programme stars are statistical errors. 
	   For $\iota$~Her systematic errors due to uncertainties of atmospheric parameters have 
	   been determined in this work and are listed in parentheses.}
\label{LTE}
\begin{tabular}{|l|r@{$\,\pm\,$}rr|r@{$\,\pm\,$}rr|r@{$\,\pm\,$}rr|r@{$\,\pm\,$}rr|} \hline
   Element 
    & \multicolumn{3}{c|}{$\iota$~Her} 
    & \multicolumn{3}{c|}{PG\,0229$+$064}
    & \multicolumn{3}{c|}{PG\,1400$+$389} 
    & \multicolumn{3}{c|}{SB\,939}\\ \hline
 $\xi$ (km $\rm{s^{-1}}$)  & \multicolumn{3}{c|}{5}      & \multicolumn{3}{c|}{5} & \multicolumn{3}{c|}{5} &  \multicolumn{3}{c|}{8} \\
 He {\sc i}    & \multicolumn{3}{c|}{10.78}  & 11.20 & 0.10 & (9)      & 11.29 & 0.10 & (6)      & 11.40 & 0.10 & (4) \\ 
 C  {\sc ii}   &  8.14 & 0.26 & ($\pm$0.21)  & 9.09  & 0.21 & (26)     &  9.08 & 0.14 & (17)     &  9.12 & 0.25 & (4) \\
 N  {\sc ii}   &  7.85 & 0.18 & ($\pm$0.14)  & 8.76  & 0.20 & (54)     &  8.89 & 0.11 & (34)     &  8.66 & 0.18 & (39) \\
 O  {\sc i}    &  \multicolumn{3}{c|}{}      & 8.81  & 0.30 &  (3)     &  8.96 & 0.06 &  (3)     & \multicolumn{3}{c|}{--} \\
 O  {\sc ii}   &  8.72 & 0.16 & ($\pm$0.19)  & 8.88  & 0.19 & (20)     &  8.89 & 0.19 & (12)     &  9.07 & 0.28 & (16) \\
 Ne {\sc i}    &  \multicolumn{3}{c|}{}      & 8.44  & 0.07 & (13)     &  8.60 & 0.09 & (12)     & \multicolumn{3}{c|}{--} \\
 Mg {\sc ii}   &  7.28 & 0.20 & ($\pm$0.12)  & 7.39  & 0.39 &  (2)     &  7.53 & 0.13 &  (2)     &  7.50 & 0.26 & (2) \\
 Al {\sc ii}   &  \multicolumn{3}{c|}{6.18}  & \multicolumn{2}{c}{6.31} & (1)     & \multicolumn{2}{c}{6.53} & (1) & \multicolumn{2}{c}{6.55} & (1)\\
 Al {\sc iii}  &  6.31 & 0.12 & ($\pm$0.14)  & 6.57  & 0.12 &  (6)     &  6.76 & 0.06 &  (5)     &  6.49 & 0.09 & (4) \\
 Si {\sc ii}   &  6.86 & 0.50 & ($\pm$0.18)  & 6.94  & 0.27 & (10)     &  7.29 & 0.27 &  (5)     &  6.66 & 0.13 & (3) \\
 Si {\sc iii}  &  7.34 & 0.20 & ($\pm$0.21)  & 7.79  & 0.13 &  (6)     &  7.93 & 0.11 &  (5)     &  7.77 & 0.24 & (4) \\
 P  {\sc ii}   &  \multicolumn{3}{c|}{}      & \multicolumn{3}{c|}{--} & \multicolumn{2}{c}{6.01}& (1)    &  \multicolumn{3}{c|}{--}  \\   
 P  {\sc iii}  &  \multicolumn{3}{c|}{5.53}  & \multicolumn{2}{c}{5.69} & (1)     & \multicolumn{3}{c|}{--} & \multicolumn{2}{c}{5.48} & (1)\\
 S  {\sc ii}   &  6.99 & 0.19 & ($\pm$0.05)  & 7.26  & 0.22 & (49)     &  7.33 & 0.19 & (35)     &  7.31 & 0.24 & (25) \\
 S  {\sc iii}  &  6.93 & 0.32 & ($\pm$0.17)  & 7.35  & 0.10 &  (2)     & \multicolumn{2}{c}{7.31} & (1) & \multicolumn{2}{c}{7.18} & (1) \\
 Ar {\sc ii}   &  6.64 & 0.30 & ($\pm$0.08)  & 7.15  & 0.10 & (15)     &  7.00 & 0.10 &  (10)    &  6.78 & 0.06 & (4) \\
 Ti {\sc ii}   & \multicolumn{3}{c|}{--}     & 5.18  & 0.15 &  (2)     & \multicolumn{3}{c|}{--} &  \multicolumn{3}{c|}{--} \\
 Fe {\sc ii}   &  7.18 & 0.23 & ($\pm$0.13)  & 7.17  & 0.13 &  (2)     &  7.31 & 0.09 &  (2)     &  7.37 & 0.16 & (2) \\
 Fe {\sc iii}  &  7.33 & 0.09 & ($\pm$0.15)  & 7.63  & 0.13 & (14)     &  7.79 & 0.15 &  (9)     &  7.49 & 0.23 & (5)\\ \hline
\end{tabular}\\
\end{table*}
\subsection{HD~76431}

The helium-poor star HD~76431 has C, N and Fe abundances quite similar 
to $\tau$~Sco. Other metals are slightly deficient with respect to
$\tau$~Sco but well within the range of main sequence B-type stars 
(except for an oxygen underabundance). 
If this star were not helium deficient it would be 
difficult to judge whether its abundance pattern is peculiar or not.

\subsection{PG\,1400$+$389 and PG\,0229$+$064}

The helium-rich stars PG\,1400$+$389 and PG\,0229$+$064 lie below the main
sequence. They display large overabundances of C (0.95\,dex) and N (0.9 to 
1.05\,dex) with respect to $\iota$~Her. While O and Ne have abundances 
similar to $\iota$~Her the other metals are somewhat more abundant than 
in the comparison stars. Such large C and N enhancement is not known to 
occur in any main sequence B-type star. Therefore it is unlikely that PG\,1400$+$389 and 
PG\,0229$+$064 are massive B-type stars.

\subsection{SB\,939}

The helium rich star SB\,939 lies on the main sequence band and might be an
intermediate helium star.
The metal abundance pattern (see Fig.~\ref{metal}) is very similar to those 
of PG\,1400$+$389 and PG\,0229$+$064. In particular, it shares the large 
enhancement of carbon and nitrogen with the latter. 
Intermediate helium stars in general display normal C, N and O abundances 
(Hunger, 1986) and therefore it appears unlikely that SB\,939 belongs to 
the class of intermediate helium stars.
\begin{table}
\centering
\caption[]{Mean absolute LTE abundances and r.m.s errors for HD~76431. The numbers in brackets indicate
           the number of lines per ion. For $\tau$~Sco the abundances are determined with the
	   equivalent widths from Hambly et al. (\cite{haro97}).
           Errors for the programme stars are statistical errors. 
	   For $\tau$~Sco systematic errors due to uncertainties of atmospheric parameters have 
	   been determined in this work and are listed in parentheses.}
\label{HD76LTE}
\begin{tabular}{|l|r@{$\,\pm\,$}rr|r@{$\,\pm\,$}rr|} \hline
   Element 
    & \multicolumn{3}{c|}{$\tau$~Sco} 
    & \multicolumn{3}{c|}{HD\,76431} \\ \hline
 $\xi$ (km $\rm{s^{-1}}$)  & \multicolumn{3}{c|}{6}      & \multicolumn{3}{c|}{5}  \\
 He {\sc i}    & \multicolumn{3}{c|}{11.10}  & 10.49 & 0.10 & (7)      \\ 
 C  {\sc ii}   & \multicolumn{3}{c|}{8.28}   & 8.49 & 0.09 & (15)      \\
 C  {\sc iii}  &  8.48 & 0.22 & ($\pm$0.14)  & 8.55 & 0.16 & (22)      \\
 N  {\sc ii}   &  8.20 & 0.19 & ($\pm$0.07)  & 8.08 & 0.17 & (65)      \\
 N  {\sc iii}  &  8.46 & 0.18 & ($\pm$0.16)  & 8.18 & 0.09 & (10)      \\
 O  {\sc ii}   &  8.60 & 0.17 & ($\pm$0.10)  & 7.81 & 0.19 & (44)      \\
 Ne {\sc ii}   &  \multicolumn{3}{c|}{}      & 8.30 & 0.17 & (4)       \\
 Mg {\sc ii}   &  \multicolumn{3}{c|}{7.53}  & 7.22 & 0.36 & (2)       \\
 Al {\sc iii}  &  6.39 & 0.03 & ($\pm$0.08)  & 5.99 & 0.24 & (5)       \\
 Si {\sc iii}  &  7.49 & 0.74 & ($\pm$0.15)  & 6.95 & 0.26 & (7)       \\
 Si {\sc iv}   &  7.50 & 0.12 & ($\pm$0.17)  & 6.72 & 0.15 & (5)       \\
 P  {\sc iii}  &  \multicolumn{3}{c|}{5.59}  & \multicolumn{2}{c}{5.04} & (1) \\
 S  {\sc iii}  &  6.87 & 0.18 & ($\pm$0.14)  & 6.59 & 0.08 & (6)       \\
 Fe {\sc iii}  &  7.36 & 0.37 & ($\pm$0.10)  & 7.22 & 0.27 & (16)       \\ \hline
\end{tabular}\\
\end{table}
\section{Evolutionary status}\label{evol_section}

Although our programme stars spectroscopically mimic normal massive B-type stars 
when studied at low spectral resolution, the high resolution spectral 
analysis presented above revealed mounting evidence that the programme
stars cannot be young massive stars. 

\begin{figure}
\vspace{8.9cm}
\includegraphics{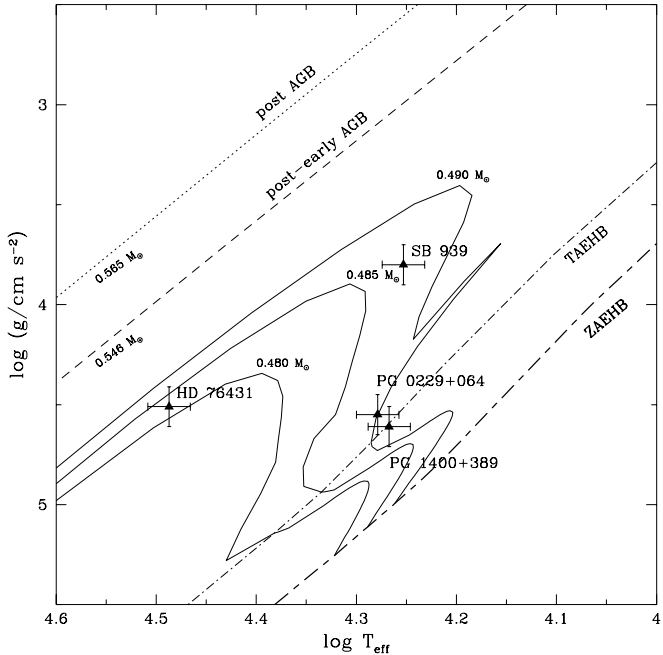}
 \caption[]{Comparison of the helium rich stars PG\,0229$+$064, PG\,1400$+$389,
            SB~939 and the helium poor star HD~76431 in the (\teff, \logg) diagram, 
            to evolutionary EHB-tracks of Dorman et al. (\cite{doro93}), 
            post-early AGB $+$ post-AGB tracks of Sch\"onberner (\cite{scho83}).
 \label{dorman}
}
\end{figure}

Low-mass stars can reach the high temperatures in question during their 
evolution in the horizontal branch, post-extreme horizontal branch 
and post-AGB phase (see Heber \cite{heb92} for a review). We therefore compare 
the positions of the four stars in the (\teff, \logg) diagram to the 
predictions of evolutionary calculations for these phases of evolution (see 
Fig.~\ref{dorman}). The gravities of all four stars are too high to 
be consistent with post-AGB evolution. PG\,0229$+$064 and PG\,1400$+$389 
lie close to the horizontal branch. However,
their abundance anomalies are unusual for horizontal branch stars which are 
mostly helium-deficient instead of helium-rich (as observed). There is 
general consensus that the abundance patterns of horizontal branch stars are 
caused by diffusion processes, which lead 
to the observed underabundances of helium due to gravitational settling.
Therefore the identification of PG\,0229$+$064 and PG\,1400$+$389 as 
horizontal branch stars may be premature.
We may speculate that the C and N enrichment may indicate that 
dredge-up of material processed by the CN cycle (N) and helium 
burning (C) may have occurred. This conjecture is corroborated by 
the observed helium enrichment. This requires deep mixing into the helium 
core in order to dredge up carbon which is unlikely to occur in horizontal 
branch stars. 

HD\,76431 and SB\,939 lie above the horizontal branch. The low 
helium abundance of HD\,76431 indicates that diffusion processes are going 
on in its atmosphere as would be expected for a star evolving from the 
horizontal branch. Comparing its position to evolutionary tracks of Dorman et 
al. (\cite{doro93}, see Fig.~\ref{dorman}) we 
conclude that HD\,76431 is in the post-EHB stage and evolves towards the 
white dwarf cooling sequence.   

The He, C and N enrichment of SB~939 again calls for a dredge-up 
mechanism which is difficult to envisage, as discussed above already.

Diffusion processes in the atmospheres of SB\,939,
PG\,0229$+$064 and PG\,1400$+$389 must be somewhat different from those in 
normal HB and post-HB stars. Physical processes that could modify diffusion
are e.g. mass loss and magnetic fields. Therefore it would be useful to search 
for the presence of magnetic fields and stellar winds in these stars.  

\section{Kinematics}\label{kine_section}

A study of their kinematics may give additional clues as to the 
nature of these stars. Proper motions for our programme stars
have become available recently (see Table \ref{properm})
through the Hipparcos/Tycho catalogs  
(Perryman et al. \cite{peli97}, H{\o}g et al. \cite{hog98}) and the work 
by Thejll et al. (\cite{thfl97}). Space motions can be derived when radial 
velocities and distance are known. These quantities can be determined 
spectroscopically.

\begin{table}
\centering
\caption[]{Data of proper motion. For the programme stars the position angle is
           counted positive east of north.}
\label{properm}
\begin{tabular}{|c|ccc|} \hline
 Name & $\mu$ (mas/y) & Position angle $^{\circ}$ & Reference \\ \hline
 PG\,0229$+$064 & 18.0 $\pm$ 3.3 & 84 $\pm$ 16 & 1 \\ 
 PG\,1400$+$389 & 6.9 $\pm$ 3.8 & 223 $\pm$ 58 & 1 \\ 
 HD\,76431      & 38.0 $\pm$ 8.0 & 233 $\pm$ 15 & 2 \\ 
 SB\,939        & 18.7 $\pm$ 15.4 & 204 $\pm$ 50 & 3 \\ \hline
\end{tabular}\\[2mm]
\begin{tabular}{ll}
\parbox[t]{150mm}{\it{References}: (1) Thejll et al. (\cite{thfl97});\\
                                   (2) Perryman et al. (\cite{peli97});
                                   (3) H{\o}g et al. (\cite{hog98})}
\end{tabular}\\
\end{table}

\subsection{Radial Velocities and Distances}
Radial velocities were derived from the lineshift of metal lines and 
corrected to heliocentric 
values. Results are listed in Table \ref{rad_dist}. 
The error of the velocities estimated 
from the scatter of the velocities derived from individual metal lines is about 
2 -- 5 km $\rm{s^{-1}}$.\\
The distance has been calculated from mass, 
effective temperature, gravity and the dereddened apparent 
magnitude of the stars:\\[2mm]
$d = 1.11 \sqrt{\frac{M_{\star}F_V}{g}\cdot 10^{0.4V_0}}$ 
[kpc]\\[2mm] where $\rm{M_{\star}}$ is the stellar mass in $\rm{M_{\odot}}$, 
g is the gravity in cm~$\rm{s^{-2}}$, $\rm{F_V}$ is the model atmosphere flux 
at the stellar surface
in units of $10^8{\,}{\rm{erg \ cm^{-2}{ }s^{-1}{ }\AA^{-1}}}$ and $\rm{V_0}$ is 
the 
dereddened apparent visual magnitude. M$_{\star}$=0.5\Msolar{  }has been 
adopted in accordance with the prediction of evolutionary models.

For HD~76431 a trigonometric parallax has been measured by Hipparcos: 
$\pi$= 3.55\,$\pm$\,1.83\,mas resulting in a distance of 280$^{+280}_{-95}$\,pc in 
good agreement with the spectroscopic distance (370$\pm$ 61 pc).

\begin{table}
\centering
\caption[]{Radial velocities and spectroscopic distances}
\label{rad_dist}
\begin{tabular}{|c|cc|} \hline
 Name & $\rm{v_{rad}}$ & d \\
      & km $\rm{s^{-1}}$  & pc  \\ \hline
 PG\,0229$+$064 & (8$\pm$2)$^\star$,(8$\pm$3)$^{\star\star}$ & 740$\pm$117 \\
 PG\,1400$+$389 & 34$\pm$2                                   & 810$\pm$134 \\
 HD\,76431      & 47$\pm$2                                   & 370$\pm$61  \\ 
 SB\,939        & $-$25$\pm$4                                & 860$\pm$134 \\ \hline
\end{tabular}\\[2mm]
\begin{tabular}{ll}
\parbox[t]{150mm} {$^\star$ obtained from KECK high resolution spectrum\\
                  $^\star$$^\star$ obtained from FOCES high resolution spectrum}
\end{tabular}\\
\end{table}

\subsection{Galactic orbits and population membership}

\begin{figure*}
\vspace{16.3cm}
 \caption[]{Meridional projections of the galactic orbits during the last 
            10\,Gyrs.
 \label{orbits}
}
\end{figure*}

In order to find out to which stellar population the stars belong we 
calculated galactic orbits backwards in time for 10\,Gyrs using the program 
ORBIT6 developed by Odenkirchen \& Brosche (\cite{odbr92}). 
This numerical code calculates the
orbit of a test body in the galactic potential of Allen \& Santillan 
(\cite{alsa91}). A detailed description of the method is given by Altmann 
and de Boer (\cite{aldb00}). The 
complete set of cylindrical coordinates is integrated and positions and
velocities are calculated in equidistant time steps. 
The input for this program version are equatorial coordinates, 
distance {\it{d}} from the sun, heliocentric
radial velocities and observed absolute proper motions. Meridional 
projections of the orbits are shown in Fig.~\ref{orbits}. The maximum 
heights above the galactic plane, the eccentricities and orbital velocities
are summarized in Table~\ref{orb_param}. These data are consistent with
HBB stars analysed by Altmann \& de Boer (\cite{aldb00}). 
The orbital velocities are slightly lower than that of $\Theta_{\rm{LSR}}$ (220\,km\,s$^{-1}$) 
and the orbits are slightly eccentric. 
The small maximum heights above the plane 
indicate that the stars belong to an 
old disk population.

\begin{table}
\centering
\caption[]{Parameters of galactic orbits z$_{max}$ is the maximum 
distance from the galactic plane, e is the eccentricity of the orbit and 
$\Theta$ is the orbital velocity.}
\label{orb_param}
\begin{tabular}{|c|ccc|} \hline
 Name           &  z$_{max}$ & e    & $\Theta$ \\ 
                & [kpc]       &                 & [km\,s$^{-1}$] \\  \hline
 PG\,0229$+$064 &  0.86       &  0.18           & 211  \\ 
 PG\,1400$+$389 &  1.29       &  0.02           & 213  \\ 
 HD\,76431      &  0.47       &  0.25           & 206  \\ 
 SB\,939        &  1.39       &  0.31           & 200  \\ \hline
\end{tabular}\\[2mm]
\end{table}

\section{Conclusions}\label{conclusions}

We have carried out differential spectral analyses of four apparently
normal B-type stars. PG\,0229$+$064, PG\,1400$+$389, HD\,76431
and SB\,939 are found to be sharp lined. This indicates that the 
projected rotational velocity is low
(v$\sin i <$5~km\,s$^{-1}$). The former three lie below the ZAMS in the (\teff,
\logg) diagram. 
Peculiar helium abundances are found,
PG\,0229$+$064, PG\,1400$+$389 and SB\,939 being helium rich, HD\,76431
being helium poor due to diffusion. Therefore we conclude that these four
stars cannot be young massive B-type stars but must be evolved low-mass 
B-type stars. 

While PG\,0229$+$064 and PG\,1400$+$389 could be explained as 
horizontal branch stars from their position in the (\teff, \logg) diagram,
HD\,76431 and SB~939 must already have evolved beyond the 
horizontal branch. The atmospheric abundance patterns of stars in 
these phases of evolution are known to be dominated by diffusion processes, 
the clear-cut indicator being a typical helium deficiency. Such an helium 
underabundance is indeed observed for HD\,76431, indicating that it 
has evolved from the extreme horizontal branch. The other stars 
are helium rich. The enrichment in helium, carbon and nitrogen can be 
explained either by 
deep mixing of nuclearly processed material to the surface or by diffusion
processes modified by magnetic fields and/or stellar winds. 
A search for magnetic fields and stellar winds is proposed.
 
Based on proper motion and radial velocity measurements and spectroscopic 
distances the galactic orbits of the stars have been calculated backwards
in time. An analysis of the orbits indicates that the stars belong to an
old disk population.
 
\acknowledgement 
{M. R. and H. E. gratefully acknowledge financial support by the DFG (grant 
He1356/27-1). We thank Neil Reid who provided us with 
the Keck observations and Thomas Rauch who obtained the
DSAZ TWIN spectra for us.}


\end{document}